\documentclass{iau}

\usepackage{amsmath}
\usepackage{graphicx}
\usepackage{multirow}
\usepackage{csquotes}

\newcommand{\rev}[1]{#1}

\begin{document}

\lefttitle{Haqq-Misra}
\righttitle{Technosignature Possibility Space}

\jnlPage{1}{7}
\jnlDoiYr{2026}
\doival{10.1017/xxxxx}

\aopheadtitle{Proceedings IAU Symposium}
\editors{J. Haqq-Misra \& R. Kopparapu, eds.}

\title{Framing the Possibility Space for Technosignature Searches}

\author{Jacob Haqq-Misra}
\affiliation{Blue Marble Space, Seattle, Washington, USA}

\begin{abstract}
This paper develops a two-parameter matrix that can be used to describe four general strategies in the search for technosignatures. The first parameter is domain accessibility: can the technosignature be accessed within the spatial domain accessible to us today? The second parameter is recognizability: would the technosignature be recognizable to us if discovered today? This yields a matrix with four options that each comprise different search strategies. ``Exploration'' is the strategy for technosignatures that are accessible within our domain and recognizable. ``Expansion'' is the strategy for technosignatures that are recognizable but beyond our spatial domain. ``Evolution'' is the strategy for technosignatures that are accessible within our domain but unrecognizable. ``Existence'' is the strategy for technosignatures that are neither within our domain nor recognizable. The implications of these four options are discussed with relevance to the Fermi paradox and strategies for searching for technosignatures.
\end{abstract}


\maketitle

\section{The Drunkard's Search}

The SETI (search for extraterestrial intelligence) discourse often invokes a metaphor to reveal the observational biases that can occur when we are constrained to only the places that are easiest to search. This idea itself is likely very old, but the modern version of this story as ``the drunkard's search'' for his lost keys appears to have first been published by Abraham Kaplan in his book \textit{The Conduct of Inquiry}:
\begin{quote}
    There is a story of a drunkard searching under a street lamp for his house key, which he had dropped some distance away. Asked why he didn't look where he had dropped it, he replied, ``It's lighter here!'' Much effort, not only in the logic of behavioral science, but also in behavioral science itself, is vitiated, in my opinion, by the principle of \textit{the drunkard's search}. \citep[][p.11]{kaplan1964conduct}
\end{quote}
This book was published just five years after the inaugural SETI paper by \cite{cocconi1959searching}. It is difficult to tell exactly when use of this metaphor entered informal SETI conversations, but it seems probable that at least some members of the early SETI community were directly or indirectly familiar with this story. 

Less than a decade after Kaplan's book, the drunkard's search was formally included in the published discussions from the first international SETI conference, convened in Soviet Armenia in 1971. The story was initially raised in the published volume by Francis Crick:
\begin{quote}
    If we want to approach the problem [of steps in evolution] in this direction rather than from the point of view of the man just looking where the light is---then you have to go at those steps which took a longish time. \citep[][p.119]{sagan1973communication}
\end{quote}
Later in the discussion, Leslie Orgel replied to this remark:
\begin{quote}
    The first point was Minsky's very much to-the-point question that if we adopt Francis Crick's very critical view, what should we do? To which Francis replied, ``You can only look where the light is; it is no good looking anywhere else.'' To which Carl Sagan in return replied that really you do have alternative strategies. \citep[][p.129]{sagan1973communication}
\end{quote}
The strategic tension between searching in the places that are easiest to look compared to the ideal locations continues to resonate in SETI community discussions today, as well as in broader conversations in the astrobiology community about where to look for biosignatures.

One way to balance between these options in SETI is to conduct searches that can also lead to ancillary results. This pragmatism remains prevalent today as echoes of the very first SETI effort by Frank Drake:
\begin{quote}
    Our experience with Project Ozma showed that the constant acquisition of nothing but negative results can be discouraging. A scientist must have some flow of positive results, or his interest flags. Thus, any project aimed at the detection of intelligent extraterrestrial life should simultaneously conduct more conventional research. Perhaps time should be divided about equally between conventional research and the intelligent signal search. From our experience, this is the arrangement most likely to produce the quickest success. \citep{1965cae..book..323D}
\end{quote}
One cannot argue with such a pragmatic approach. Indeed, Kaplan even noted that the drunkard's strategy of searching under the light may be a reasonable place to begin:
\begin{quote}
    The drunkard's search is relevant here; the pattern of search, we feel, should be closely related to the probability of the thing sought being in the place where the seeker is looking. But the joke may be on us. It may be sensible to look first in an unlikely place just \textit{because} ``it's light there''. We might reasonably entertain one hypothesis rather than another because it is easier to refute if false, or because it will eliminate a greater number of possibilities, or because it will show us more clearly what steps to take next. \citep[][p.17]{kaplan1964conduct}
\end{quote}

\rev{In the search for extraterrestrial life, the reporting of negative results can provide important information about the limits of habitability} \citep{ratliff2023vacant}. Given the variety of approaches that a search for keys in the dark (or technosignatures in space) can encompass, this paper attempts to utilize the metaphor of the drunkard's search to explore the ``alternative strategies'' (as Sagan remarked) or possibility space for generalizing the available options.

\section{Accessibility and Recognizability}\label{sec:access}

\rev{This analysis of the drunkard's search, and its analogical mapping to the search for technosignatures, begins with two assumptions:}
\begin{quote}
    \rev{A1: The drunkard does not know where to look for the keys.}\\\\
    \rev{A2: The drunkard knows the keys exist.}
\end{quote}
\rev{In this section, the ontological reality of the keys is not under question. The drunkard is unsure of where to search, but he does know that he is searching for something real. In the search for technosignatures, the first assumption (A1) appropriately can be modified as ``scientists do not know where to look for technosignatures.'' However, the second assumption (A2) does not correctly correspond to the search for technosignatures, as scientists searching for technosignatures are not even sure if any exist. We will retain the assumptions A1 and A2 in this section for now, and will revisit them again in \S\ref{sec:fermi}.}

\rev{The \textit{spatial domain} is defined as the physical area that can be searched with available physical senses and remote sensing tools. For the drunkard, the spatial domain is the area illuminated by the light.} The foremost question of the drunkard's search is: are the keys \rev{located} in the illuminated spatial domain? If the answer is yes, then searching under the light is a reasonable strategy, and the drunkard will hopefully soon find his keys. But if the answer is no, then searching under the light is a fruitless endeavor, and a better strategy would involve obtaining a portable light to illuminate a new area \rev{(i.e., expand the spatial domain)} to continue the search. This aspect of the problem is described as the parameter \textit{accessibility}: \rev{the keys are \textit{accessible} if they are located inside the spatial domain (i.e., the illuminated area), and the keys are \textit{inaccessible} if they are located outside the spatial domain (i.e., the dark region beyond the light).} 

A second parameter of the problem is \textit{recognizability}: \rev{the keys are \textit{recognizable} if the drunkard would be able to find them using his available physical senses and remote sensing tools, and the keys are \textit{unrecognizable} if the drunkard would not be able to find them using his available physical senses and remote sensing tools. For the drunkard, eyesight is the primary physical sense, aided by the reflected light provided by the lamp.} If the drunkard would be able to recognize the keys if he encountered them, then \rev{he should simply continue searching within his spatial domain (the illuminated area) until he eventually finds the keys (although it may take a long time)}. If instead the keys are not recognizable but still within the spatial domain \rev{(perhaps concealed in a mound of dirt or leaves)} then the drunkard may need to \rev{expand his sensory capabilities (e.g,} obtain new tools, such as a metal detector) in order to be successful.  But if the keys are not recognizable and outside the spatial domain \rev{(i.e., located beyond the light and concealed from vision)}, then the drunkard's only option is to wait for somebody else to find the keys and bring them to him. 

\begin{figure}[!t]
    \centering
     \includegraphics[width=7cm]{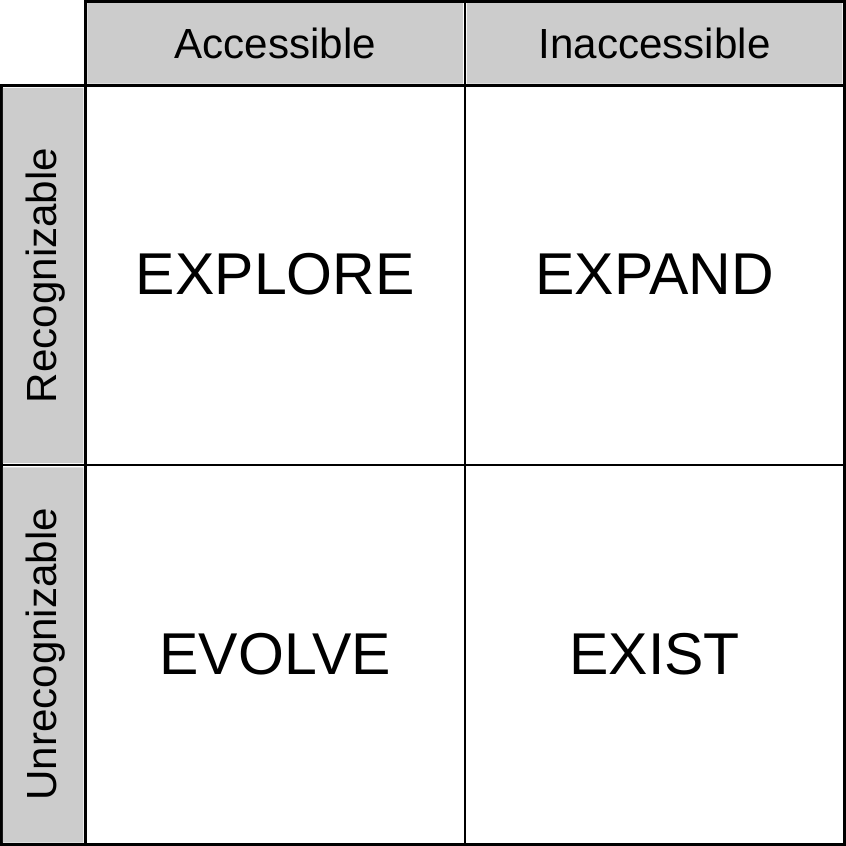}
     \caption{The two parameters of accessibility and recognizability give four strategic options in the drunkard's search.} 
    \label{fig:matrix}
\end{figure}

In the search for technosignatures, the parameters of this metaphor map onto two questions. First, is the technosignature in our spatial domain? \rev{Here the spatial domain is the physical area of the universe that scientists are able to search using remote sensing tools. The spatial domain includes Earth and the solar system plus all known exoplanets, stars, and galaxies; the spatial domain excludes anything beyond these bounds that cannot presently be observed, such as as-yet unknown exoplanets and as-yet undiscovered high-redshifted galaxies.} Second, would we recognize the technosignature if encountered? \rev{Here a technosignature is recognizable if we would be able to detect it using available remote sensing tools, which includes acknowledgment of any detection limits that may exist. For example, if the industrial pollutants CFC-11 and CFC-12 are abundant in the atmosphere of TRAPPIST-1 e, then these technosignature candidates could be detected through transit spectroscopy by the James Webb Space Telescope \citep{haqq2022detectability}, thus being recognizable. However, if TRAPPIST-1 e hosted a belt of orbiting satellites similar to Earth today, then this would be undetectable by any current or foreseeable observatory \citep{sallmen2019improved}, thus being unrecognizable.} This possibility space is visually summarized in Figure \ref{fig:matrix}. 

If the answer to both questions is yes, then the solution is to ``Explore'' in the sense of continuing to search and exploit available resources to find any technosignatures that we have not yet noticed. \rev{This is the most optimistic case in which a technosignature exists within our spatial domain and could conceivably be recognized with our current observatories: in such a scenario, we simply need to continue searching with our current tools, and we will eventually succeed (although the search may still take a long time). For example, the search space that has been covered by radio SETI surveys is analogous to the ratio of the volume of water in a small swimming pool compared to the volume of Earth's oceans \citep{wright2018much}, and it remains possible that future radio searches will find promising technosignature candidates.}

If a technosignature is not in our spatial domain but would be recognizable if encountered, then the strategy is to ``Expand' the size of our spatial domain to encompass a greater volume of targets to search. \rev{This includes future developments that expand our abilities to find and characterize known targets. For example, the Habitable Worlds Observatory is a mission under study at NASA Goddard Space Flight Center that would enable reflected light detection and characterization of Earth-like planets orbiting Sun-like stars, which could enable searches for technosignature candidates such as nitrogen dioxide in exoplanet atmospheres \citep{kopparapu2021nitrogen}. Likewise, the Square Kilometre Array under construction in Australia and South Africa will significantly increase the sensitivity of radio surveys to possible technosignatures \citep{siemion2014searching}.}

A technosignature within our spatial domain but that is unrecognizable would challenge us to ``Evolve'' or advance our search to new sensory modalities \rev{beyond our current capabilities. This encompasses the possibility that evidence of extraterrestrial technology exists in targets that we already know about (such as solar system planets or known exoplanetary systems), but we do not possesses the technological or cognitive ability to find it. Examples include the idea that ``postbiological evolution'' could lead to technospheres completely unrecognizable to us \citep[e.g.,][]{cirkovic2006galactic} or that a long-duration extraterrestrial technosphere may share no detectable properties in common with Earth's present-day technosphere \citep[e.g.,][]{garrett2025blink}. Addressing this challenge would require new theoretical ideas about \textit{what to look for} and instrumentation for conducting the search.}

Finally, a technosignature that is neither within our spatial domain nor recognizable would require us to ``Exist'' or wait in hope that evidence of extraterrestrial technology somehow finds us. \rev{Another option could be to simultaneously expand and evolve, in the hope of finding new targets and new modalities of characterizing them. This may lead to other astrophysical discoveries; however, this also amounts to a strategy of poking around in the dark, lacking any guidance for where and how to search.}

\section{Search Strategies and the Fermi Paradox}\label{sec:fermi}

This dissection of the drunkard's search metaphor can be useful for thinking categorically about the possibility space for technsignature searches. \rev{However, the framing of the problem in \S\ref{sec:access} assumed that the keys (technosignatures) are known to exist (assumption A2). In this section, the assumptions are restated as:}
\begin{quote}
    \rev{B1: Scientists do not know where to look for technosignatures.}\\\\
    \rev{B2: Scientists do not know if any technosignatures exist.}
\end{quote}
\rev{If a scientist (or the drunkard) is unsure if there is even an object to look for (B2), then the search strategy depends on the \textit{prior odds} of one strategy over another.} 

\rev{Determining the prior odds for expecting a technosignature in a given search space is a subjective probabilistic assessment, which may be informed by data but still can lead to choices of different values by different people \citep[e.g.,][]{haqq2012likelihood}. The different prior odds estimated by individual scientists will lead each toward different strategic options in the search. For example,  some scientists may interpret the current set of available data to conclude that the best chance of finding technosignatures is to continue looking for radio signals; others may prefer prioritizing large space telescopes for exoplanet atmospheric characterization; others may argue that searching the solar system for active or defunct extraterrestrial probes is the most likely to succeed. The preference for one particular search strategy over another also correlates to attitudes regarding the Fermi paradox.} These options are summarized in Table \ref{tab:options}. 

The strategic option of ``Explore'' includes looking for radio, optical, and other recognizable signals that are presently observable at Earth, looking for artifacts (active or defunct) that may be residing in the solar system, \rev{and searching known exoplanets for atmospheric technosignatures with current observatories. Within an environment of limited resources and competitive funding, one way of prioritizing specific exploratory searches for technosignatures is to use a framework such as the ``Axes of Merit'' \citep{sheikh2020nine} in order to balance to expectations of a technosignature discovery with other factors such as cost and ancillary benefits.} 

The strategy to ``Expand'' includes finding new targets to observe, including exoplanets and perhaps other locations that may make sense for thinking broadly about technosignature possibilities. Expansion includes remote observations but also could take the form of in situ observations with robotic or crewed missions. Expansion likewise includes extending the general set of astronomical observations. This strategy includes near-term expansion to other planets in the solar system as well as longer-term ambitions for expansion of the spatial domain to interstellar scales. \rev{Using the Axes of Merit \citep{sheikh2020nine}, this strategy would apply to any technosignature that ranks very low on the ``Observing Capability'' axis, meaning that far-future assets would be needed to detect the technosignature.}

\begin{table}
    \centering
    \begin{tabular}{p{0.1\linewidth}|p{0.4\linewidth}p{0.4\linewidth}}\toprule
         \rev{Option} &  Fermi Paradox \rev{Prior Odds} & Technosignature Search Strategy \\\midrule
         Explore & \rev{They likely} \rev{exist, and} we will find them if we keep looking. &  Look for radio, optical, and other signals. Look for artifacts in the solar system. \rev{Characterize exoplanet atmospheres.}\\ \\ \\
         Expand & They \rev{likely exist}, \rev{but} we have to expand if we want to find them. &  Find new targets to observe. Visit new targets with robotic and crewed missions. Develop new capabilities to search known systems.\\ \\ \\ 
         Evolve & They \rev{likely exist}, but we would not recognize them if we found them. &  Develop new \rev{theoretical ideas about what to look for and instrumentation for conducting the search.}\\ \\ \\
         Exist & They \rev{may or may not exist, and} we cannot find them \rev{if they do exist}. &  Patience and serendipity.\\ \\ \bottomrule
    \end{tabular}
    \caption{The four strategic options of the drunkard's search \rev{imply different assumed prior odds } for the Fermi paradox, \rev{which correspond to different preferences} for technosignature searches.}
    \label{tab:options}
\end{table}

The strategy to ``Evolve'' does not necessarily require us to wait for biological evolution to solve the problem, as we can advance our search capabilities by developing new \rev{theoretical ideas and instrumentation}. Such improved approaches would enable novel searches within our spatial domain, which may include re-examination of previously searched systems, and could reveal anomalies that would not otherwise be detected. \rev{Using the Axes of Merit \citep{sheikh2020nine}, this strategy would apply to any technosignature that ranks very low on the ``Ambiguity'' and ``Extrapolation'' axes, meaning that the technosignature may have little resemblence with Earth technology and may be easily misidentified.}

Finally, the strategy to ``Exist'' gives no options other than to stay patient, \rev{perhaps poking around blindly in the dark,} and hope that serendipity comes our way. \rev{Using the Axes of Merit \citep{sheikh2020nine}, this strategy would apply to any technosignature that ranks very low on the ``Detectability'' axis, meaning that it is inaccessable to any of our modes of searching.}

These search strategies correspond to \rev{different assumed prior odds} for the Fermi paradox (Table \ref{tab:options}). If ``Explore'' is the \rev{assumed} solution, then this implies that extraterrestrial technology is here, either from extant extraterrestrial technological activity within our observing capabilities or artifacts of previous technological activity by an extraterrestrial society that has now gone extinct or moved on. This likewise implies that we should keep searching with our existing methods until we find technosignatures. This \rev{assumption} corresponds to the ``They Are (or Were) Here'' category of Fermi paradox solutions tabulated by \citet{webb2015if}.

\rev{If} ``Expand'' \rev{is the assumed solution, then this} implies that no extraterrestrial technological activity has yet reached Earth, either past or present. This does not mean they are nonexistent, but it does mean that any process of galactic-scale settlement that may have occurred was sufficiently incomplete (for whatever reason) to omit Earth. \rev{If} ``Evolve'' \rev{is the assumed solution, then this} implies that extraterrestrial technology, or its relics, do exist in our spatial domain, but we will not recognize it with our current perceptual capabilities. In this case, galactic-scale settlement may have successfully occurred in the way posed by the Fermi paradox, and we simply are unaware that the solar system has been settled. Both of these \rev{assumptions} are clustered together in the ``They Exist, But We Have Yet to See or Hear from Them'' category by \citet{webb2015if}.

\rev{If} ``Exist''  \rev{is the assumed solution, then this implies nothing about whether or not extraterrestrial being exist. This} includes the Fermi paradox resolution that we are alone as a technological society. But this also includes other possibilities, such as technology being rare enough that contact between technospheres is non-existent, or the asymmetry between other technospheres and our own is sufficiently vast that any efforts at searching, expanding, and advancing will be in vain on our part. This \rev{assumption} includes the ``They Don't Exist'' category by \citet{webb2015if} as well as a handful of other Fermi paradox resolutions.

\section{Conclusion}

It is impossible to know which of the four strategies---Explore, Expand, Evolve, or Exist---is likely to find technosignatures. \rev{Individual preferences for any of these strategies reveal underlying assumptions about an individual's prior odds, which correspond to different assumptions about the Fermi paradox.} 

\rev{Relying on one's own subjective prior odds to guide the search for technosignatures may or may not be the best strategy. Another approach could be} to develop theoretical models to choose between competing strategies, which may be useful in situations when resources are limited. Yet another approach is to pursue all strategies simultaneously, as technology and resources permit, which would include low-cost commensal observations, dedicated searches when possible, and creative thinking about new technosignatures as time permits. Indeed, this describes how the technosignature community has advanced, and continually revisiting our core assumptions about what is accessible and what is recognizable may help us to eventually find the object of our search. If all one is able to do (or is paid to do) is search under the light, then we should nevertheless keep looking.

\section*{Acknowledgments}

\noindent The author asked the Claude large language model for a reference to the oldest instance of the drunkard's search in the SETI literature, but Claude could not provide an answer and instead gave Jill Tarter as a reference. The author then asked Jill Tarter, who agreed that the citations in this paper are probably the oldest and who was also pleased to have been considered an expert by an AI. \rev{Thanks also to Ravi Kopparapu for helpful comments that improved this paper.} The author gratefully acknowledges support from the NASA Habitable Worlds program under grant 80NSSC24K1896. Any opinions, findings, and conclusions or recommendations expressed in this material are those of the author and do not necessarily reflect the views of any employer or NASA.


\begin{thebibliography}{}

\bibitem[{\'C}irkovi{\'c} and Bradbury, 2006]{cirkovic2006galactic}
{\'C}irkovi{\'c}, M.~M. \& Bradbury, R.~J. 2006,
\newblock Galactic gradients, postbiological evolution and the apparent failure of SETI. {\em New Astronomy}, 11(8), 628--639.

\bibitem[Cocconi and Morrison, 1959]{cocconi1959searching}
Cocconi, G. \& Morrison, P. 1959,
\newblock Searching for interstellar communications. {\em Nature}, 184(4690), 844--846.

\bibitem[{Drake}, 1965]{1965cae..book..323D}
{Drake}, F.~D. 1965,
\newblock {The Radio Search for Intelligent Extraterrestrial Life}.
\newblock In {Mamikunian}, G. \& {Briggs}, M.~H., editors, In {\em Current Aspects of Exobiology}, pp. 323--345. Oxford University Press.

\bibitem[Garrett, 2025]{garrett2025blink}
Garrett, M.~A. 2025,
\newblock Blink and you’ll miss it - How technological acceleration shrinks SETI’s narrow detection window. {\em Acta Astronautica}, 238, 160--167.

\bibitem[Haqq-Misra and Kopparapu, 2012]{haqq2012likelihood}
Haqq-Misra, J., \& Kopparapu, K. 2012,
\newblock On the likelihood of non-terrestrial artifacts in the solar system. {\em Acta Astronautica}, 72, 15--20.

\bibitem[Haqq-Misra et al., 2022]{haqq2022detectability}
Haqq-Misra, J., Kopparapu, K., Fauchez, T.~J., Frank, A., Wright, J.~T. \& Limgam, M. 2022,
\newblock Detectability of chlorofluorocarbons in the atmospheres of habitable M-dwarf planets. {\em The Planetary Science Journal}, 3(3), 60.

\bibitem[Kaplan, 1964]{kaplan1964conduct}
Kaplan, A. 1964,
\newblock {\em The Conduct of Inquiry: Methodology for Behavioral Science}.
\newblock Chandler Publishing Company.

\bibitem[Kopparapu et al., 2021]{kopparapu2021nitrogen}
Kopparapu, R., Arney, G., Haqq-Misra, J., Lustig-Yaeger, J. \& Villanueva, G. 2021
\newblock Nitrogen dioxide pollution as a signature of extraterrestrial technology. {\em The Astrophysical Journal}, 908(2), 164.

\bibitem[Ratliff et al., 2023]{ratliff2023vacant}
Ratliff, L.~E., Fulford, A.~H., Pozarycki, C.~I., Wimp, G., Nichols, F., Osburn, M.~R., Graham, H.~V. 2023,
\newblock The vacant niche revisited: Using negative results to refine the limits of habitability.  {\em bioRxiv}, doi:10.1101/2023.11.06.565904.

\bibitem[Sagan, 1973]{sagan1973communication}
Sagan, C. 1973,
\newblock {\em Communication with Extraterrestrial Intelligence (CETI)}.
\newblock MIT Press.

\bibitem[Sallmen et al., 2019]{sallmen2019improved}
Sallmen, S., Korpela, E. \& Crawford-Taylor, K. 2019,
\newblock Improved analysis of Clarke exobelt detectability. {\em The Astronomical Journal}, 158(6), 258.

\bibitem[Sheikh, 2020]{sheikh2020nine}
Sheikh, S.~Z. 2020,
\newblock Nine axes of merit for technosignature searches. {\em International Journal of Astrobiology}, 19(3), 237--243.

\bibitem[Siemion et al., 2014]{siemion2014searching}
Siemion, A.~P.~V., Benford, J., Cheng-Jin, J., Chennamangalam, J., Cordes, J., DeBoer, D.~R., Falcke, H., Garrett, M., Garrington, S., Gurvits, L., et al. 2014,
\newblock Searching for extraterrestrial intelligence with the Square Kilometre Array.  {\em arXiv}, doi:10.48550/arXiv.1412.4867.

\bibitem[Webb, 2015]{webb2015if}
Webb, S. 2015,
\newblock {\em If the Universe is Teeming with Aliens... Where Is Everybody?: Seventy-five Solutions to the Fermi Paradox and the Problem of Extraterrestrial Life}.
\newblock Springer.

\bibitem[Wright et al., 2018]{wright2018much}
Wright, J.~T., Kanodia, S. \& Lubar, E. 2018,
\newblock How much SETI has been done? Finding needles in the n-dimensional cosmic haystack. {\em The Astronomical Journal}, 156(6), 260.

\end{thebibliography}

\end{document}